\documentclass{vgtc}                          

\graphicspath{{figures/}{pictures/}{images/}{./}} 
\usepackage{enumitem}
\usepackage{times}                     

\usepackage{tabu}                      
\usepackage{booktabs}                  
\usepackage{lipsum}                    
\usepackage{mwe}                       

\usepackage{mathptmx}                  

\usepackage{amsmath,amssymb,amsfonts}
\usepackage{multicol}
\usepackage{multirow} 
\usepackage{algorithm}
\usepackage{algpseudocode}

\onlineid{0}

\vgtccategory{Research}

\vgtcinsertpkg

\title{Coordinated 2D-3D Visualization of Volumetric Medical Data \\in XR with Multimodal Interactions}

\author{Qixuan Liu\thanks{Email: qixuanliu@link.cuhk.edu.hk}$^{\;\; 1,2}$, Shi Qiu\thanks{Corresponding author: shiqiu@cse.cuhk.edu.hk}$^{\;\; 1,2}$, Yinqiao Wang$^{1,2}$, Xiwen Wu$^{1,2,5}$, \\Kenneth Siu Ho Chok$^{2,3,4}$, Chi-Wing Fu$^{1,2}$, Pheng-Ann Heng$^{1,2}$\\ %
        \scriptsize $^{1}$Department of Computer Science and Engineering, The Chinese University of Hong Kong\\
        \scriptsize $^{2}$Institute of Medical Intelligence and XR, The Chinese University of Hong Kong\\
        \scriptsize $^{3}$Department of Surgery, The Chinese University of Hong Kong\\
        \scriptsize $^{4}$Prince of Wales Hospital, Hong Kong\\
        \scriptsize $^{5}$Zhujiang Hospital, Southern Medical University}

\abstract{
Volumetric medical imaging technologies produce detailed 3D representations of anatomical structures. However, effective medical data visualization and exploration pose significant challenges, especially for individuals with limited medical expertise. We introduce a novel XR-based system with two key innovations: (1) a coordinated visualization module integrating Multi-layered Multi-planar Reconstruction with 3D mesh models and (2) a multimodal interaction framework combining hand gestures with LLM-enabled voice commands. We conduct preliminary evaluations, including a 15-participant user study and expert interviews, to demonstrate the system's abilities to enhance spatial understanding and reduce cognitive load. Experimental results show notable improvements in task completion times, usability metrics, and interaction effectiveness enhanced by LLM-driven voice control. While identifying areas for future refinement, our findings highlight the potential of this immersive visualization system to advance medical training and clinical practice. Our demo application and supplemental materials are available for download at: \url{https://osf.io/bpjq5/}.
}

\begin{document}



\maketitle

\section{Introduction}
\label{sec:intro}


Volumetric medical imaging technologies, including Computed Tomography (CT) and Magnetic Resonance Imaging (MRI), have transformed modern healthcare by providing detailed 3D anatomical representations that are critical for diagnosis and surgical planning~\cite{Paudyal2023Artificial}. Despite their clinical value, effectively visualizing and exploring these modalities of volumetric medical data remains challenging, particularly for non-specialists lacking medical training.

\begin{figure}[t]
    \centering
    \includegraphics[width=.9\linewidth]{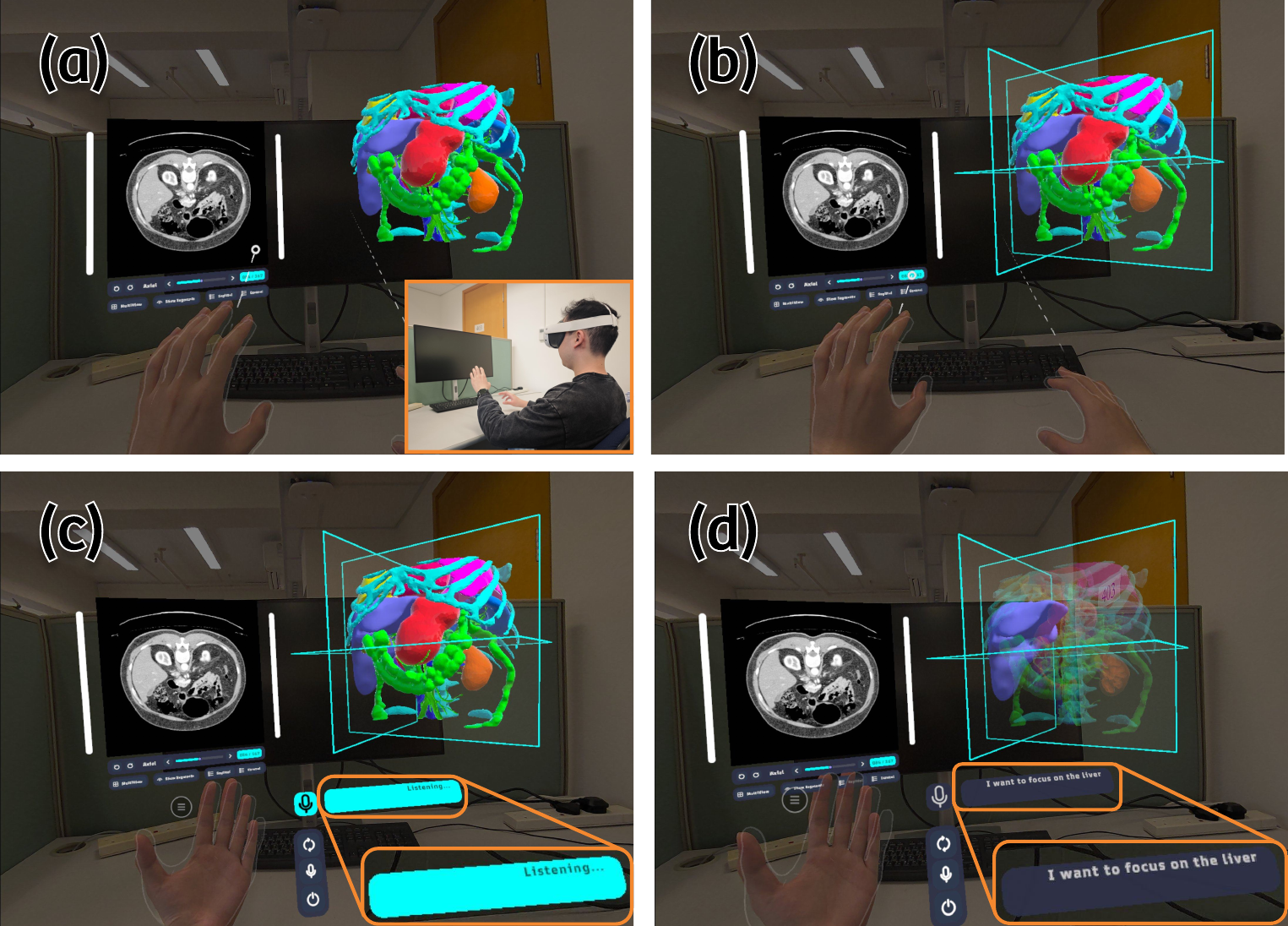}
    \caption{Overview of the user interfaces. (a) The default user interface without coordination. (b) The user interface with coordination using Volumetric Slice Planes (VSPs). (c) The user interface when using voice commands. (d) The visual result given by a voice command of ``\emph{I want to focus on the liver}''.}
    \label{fig:system_overview}
    \vspace{-3mm}
\end{figure}

Current clinical workflows mainly rely on 2D image slices to conceptualize 3D spatial relationships between anatomical structures. This process demands not only extensive training but also exceptional spatial reasoning skills, causing a steep learning curve for early-stage practitioners and non-expert users \cite{guillot2007relationship}. While researchers have developed advanced visualization techniques like volume rendering, 3D reconstruction, and 3D printing to address these limitations, 
persistent barriers are noticed: resource-intensive computations restrict real-time volume rendering~\cite{engel2000combining}, 2D displays of 3D models inadequately convey depth perception~\cite{byl2020moving}, and 3D-printed physical models lack dynamic interactivity~\cite{martelli2016advantages}.


Extended Reality (XR) has recently emerged as a powerful tool for bridging the gap between static 2D radiological images and 3D anatomical models \cite{huang2021phoenix}. By offering immersive environments with real-time passthrough capabilities and interactive 3D manipulation, XR systems enable users to explore complex anatomical structures with enhanced spatial awareness. However, some issues still exist in practice. For instance, users often face difficulties correlating 2D images with 3D models, which can lead to cognitive overload and reduced spatial comprehension \cite{garcia2024immersive}. Also, designing a natural and efficient interaction mechanism for volumetric data manipulation introduces systematic complexities in XR \cite{mandalika2018hybrid,10896112}.

\begin{figure*}[t]
    \centering
    \includegraphics[width=0.85\linewidth]{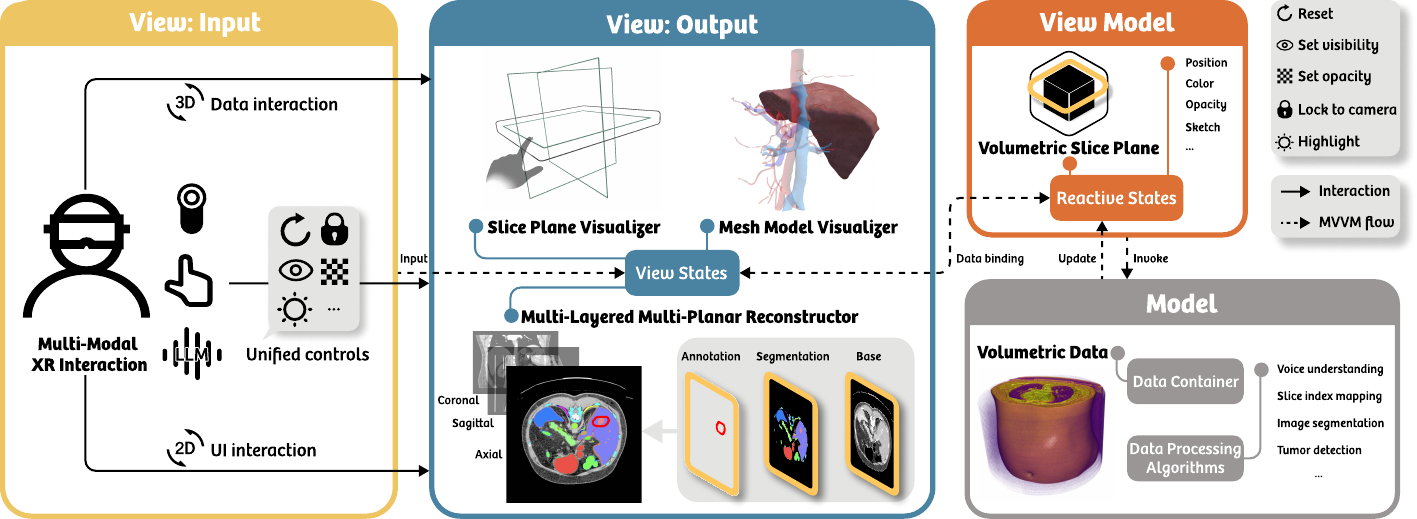}
    \caption{The pipeline of our multimodal coordinated 2D-3D visualization system.
    }
    \label{fig:pipeline}
\end{figure*}
To address these challenges, we present a new XR-based system for volumetric medical data visualization, devising two key innovations: \textbf{(1) a coordinated visualization module} that integrates Multi-layered Multi-planar Reconstruction (MLMPR) with 3D mesh models, providing synchronized and interactive views of 2D radiological slices alongside 3D anatomical structures; and \textbf{(2) a multimodal interaction framework} that combines intuitive hand gestures with voice commands powered by Large Language Models (LLMs), enabling natural, efficient, and customizable data exploration. Moreover, we conduct a comprehensive evaluation of the developed system, demonstrating its potential to enhance spatial understanding, improve usability, and reduce cognitive load, particularly for users with limited medical training.
\section{Related Work}
\noindent\textbf{Medical Data Visualization.}
To interpret complex 3D anatomical structures, volume rendering \cite{max1995optical} provides rich visualizations of subtle structural details but require significant computational resources \cite{heng2006gpu}. 3D reconstruction methods \cite{khan2018methodological} can improve interactivity and comprehensibility but may omit critical details during the reconstruction process. While 3D printing provides tangible models for hands-on exploration \cite{yan2023enhancing}, its static physical nature limits flexibility for real-time modifications. Recent XR technologies devise immersive 3D environments \cite{klonig2020integrating} for medical data visualization \cite{venkatesan2021virtual,qiu2025cvhslicer}. Particularly, integrating XR with 3D reconstructed models allows users to explore intricate anatomical structures more naturally, benefiting educational and clinical applications \cite{garcia2024immersive}. 

\noindent\textbf{Multimodal Interaction in XR.}
Integrating multiple interaction modalities in XR environments not only enhances usability but also improves task performance \cite{chikobava2023multimodal}. Friedrich \emph{et al.}~\cite{friedrich2021combining} showed that combining hand gestures with voice commands in VR led to more efficient object manipulation and reduced cognitive load. Similarly, Shi \emph{et al.}~\cite{shi2023exploring} found that combining eye-tracking with hand-based interactions improved selection accuracy in 3D spaces. Recently, Heinrich \emph{et al.}~\cite{HEINRICH2023103117} applied keyword recognition to realize voice-based interactions, showing feasibility but with limited flexibility. Our system leverages multimodal interaction for medical data visualization, through the integration of LLMs with XR interfaces. 
By combining hand gestures with LLM-enabled voice commands in a unified framework, we provide intuitive, efficient, and natural control over medical visualization tasks.

\section{System Design}
Our XR-based system aims to enhance volumetric medical data exploration by integrating coordinated 2D-3D visualization and multimodal interaction. Fig.~\ref{fig:pipeline} illustrates the system pipeline, which consists of two fundamental functional parts: (1) a coordinated visualization module that synchronizes 2D radiological slices with 3D anatomical models using a spatial coherence algorithm; and (2) a multi-modal interaction framework that combines intuitive hand gestures with LLM-enabled voice commands for seamless control of visualization tasks. Both components operate within a unified coordinate system, ensuring consistent spatial relationships and intuitive interactions across all visualization modes.

\subsection{Coordinated 2D-3D Visualization}
This module bridges the gap between 2D radiological data and 3D anatomical representations, addressing the challenge of maintaining spatial coherence in XR environments. By integrating \emph{Multi-layered Multi-planar Reconstruction (MLMPR)} with interactive 3D mesh models, users can simultaneously explore detailed 2D slices and 3D structures, facilitating a comprehensive understanding of volumetric medical data.

\noindent\textbf{2D MLMPR Visualization.}
The MLMPR component implements a multi-layered rendering pipeline designed primarily for visualizing 2D slices of volumetric medical data. Additionally, it provides dynamic overlay mechanisms to control the visualization of segmentation results and user annotations, enabling real-time updates and context-rich exploration. By structuring visualization into distinct layers, MLMPR allows users to flexibly toggle overlays without disrupting the core slice visualization, thus preserving clear logical relationships between visual elements.

To define slice orientations within the 3D volume, we introduce Volumetric Slice Planes (VSPs). The system implements three orthogonal VSPs aligned with the model's principal axes, corresponding to the standard anatomical planes: sagittal (YZ), coronal (XZ), and axial (XY). Each VSP is characterized by a position vector $\mathbf{p} \in \mathbb{R}^3$, orientation normal $\mathbf{n} \in \mathbb{R}^3$ aligned with the respective anatomical axis, and viewing dimensions $\mathbf{s} \in \mathbb{R}^2$. This orthogonal arrangement ensures consistency with radiological conventions, allowing users to navigate the dataset using familiar medical visualization paradigms.

\noindent\textbf{3D Mesh Visualization.}
We use polygon meshes for 3D visualization due to their optimized efficiency in XR applications~\cite{sutherland_applying_2019} and proven effectiveness in clinical diagnosis~\cite{byl2020moving}. Importantly, the detailed structures captured by 3D meshes offer clinicians precise views of anatomical features, crucial anatomical landmarks, and surgical targets~\cite{huang2021phoenix,zhu2024ssp}. In practice, the interactive nature of XR allows clinicians to selectively visualize specific anatomical regions, reducing cognitive load and enhancing diagnostic effectiveness.

Our system utilizes either predefined mesh models generated from volumetric segmentation data~\cite{cui2023adaptive} or manually created models. A key requirement is that anatomical structures shall be pre-segmented to establish correspondences between 2D and 3D contexts. To ensure seamless correlation and preserve visual consistency between 2D and 3D views, we develop a robust mapping scheme between mesh models and their corresponding segmentation overlays, aiding users in intuitively exploring anatomical structures across different visualization modes.

\noindent\textbf{Spatial Coherence.}
We develop a bidirectional spatial coherence algorithm to establish precise mapping between 3D world coordinates and volume voxel indices. This algorithm is crucial for maintaining correspondence between VSPs' positions in 3D space and their respective slice indices in the volume data, realizing synchronized control through either 2D UI interactions or direct 3D VSP manipulation. The core algorithm implements two key bidirectional mappings: (1) \emph{World-to-Volume mapping} ($f_{w2v}: \mathbb{R}^3 \rightarrow \mathbb{Z}^3$) that transforms world-space coordinates to volume slice indices, and (2) \emph{Volume-to-World mapping} ($f_{v2w}: \mathbb{Z}^3 \rightarrow \mathbb{R}^3$) that converts volume indices back to world coordinates. These mappings enable seamless transitions between 2D slice and 3D spatial navigations.

To accommodate meshes with varying sizes and coordinate systems, we introduce an \emph{automatic calibration} process that establishes spatial alignment through boundary matching. The calibration automatically determines scaling factors and offset parameters by analyzing the bounding boxes of both the segmentation volume (in voxel space) and the mesh model (in local coordinate space). This process maps the mesh's local origin to the corresponding position in volume space using percentage-based interpolation, ensuring accurate coordinate correspondence regardless of the mesh model's origin or scale.

Once calibrated, the derived mapping functions enable accurate spatial coherence across all visualization modes, supporting reliable anatomical interpretation in both 2D and 3D contexts. The algorithmic details are provided in the supplemental material.

\noindent\textbf{View Synchronization.}
To maintain coherence across all views, our system adopts a reactive MVVM architecture where all reactive properties are centralized in the ViewModel layer, as illustrated by the dashed arrows in Fig.~\ref{fig:pipeline}. The ViewModel serves as a reactive hub that automatically propagates state changes between any components through an \emph{any-to-any state notification} mechanism. This approach ensures that modifications from any interaction source (2D UI, 3D manipulation, voice commands) trigger appropriate coordinate transformations via $f_{w2v}$ or $f_{v2w}$ and immediately reflect across all visualization modes. The complete synchronization procedures are detailed in the supplementary material.

Overall, our coordinated visualization module achieves the following objectives: (1) enhancing structural correlation between 2D radiological slices and 3D anatomical models; (2) facilitating spatial cognition of complex anatomical relationships through synchronized visualization; and (3) reducing cognitive load during volumetric data exploration by enabling seamless transitions between 2D and 3D representations.

\subsection{Multi-modal Interaction}
To complement the visualization module, our system applies a multimodal interaction framework that combines intuitive hand gestures with the advanced LLM-enabled voice commands. This framework facilitates an easier exploration of volumetric medical data, catering to diverse preferences and clinical scenarios.

\noindent\textbf{Hand-based Interaction.}
Our system supports a range of natural hand gestures for direct manipulation of both 2D and 3D content, providing intuitive control over regular visualization operations:

\begin{itemize}[noitemsep,topsep=0pt]
    \item \textbf{Pinch:} For object selection and manipulation.
    \item \textbf{Open palm:} To activate menus or reset views.
    \item \textbf{Raycasting:} For distant object selection and interaction.
    \item \textbf{Poke:} For direct and precise UI interactions, such as toggling overlays or selecting menu options.
\end{itemize}

The designed gestures allow users to perform tasks such as rotating, scaling, translating 3D models, and navigating through 2D slice stacks. Using these familiar user actions, we aim to minimize learning curve and enhance user engagement.

\noindent\textbf{LLM-enabled Voice Interaction.}
Initially, speech signals are converted to text via a robust speech recognition API. The converted text then undergoes LLM-powered intent recognition and entity detection, allowing the system to understand complex medical terminology and contextual nuances. Finally, the recognized intents are translated into specific visualization operations through a dedicated action execution module. In practice, our system supports two categories of voice commands: (1) \emph{Basic navigation commands} that enable users to control visualization parameters directly, such as switching views or navigating slice positions (\emph{e.g.}, ``Switch to sagittal view'', ``Move to slice 99''); and (2) \emph{Advanced commands} that allow users to combine multiple actions into a single command, such as highlighting specific anatomical structures or adjusting visualization settings (\emph{e.g.}, ``Show the liver'', ``Highlight the tumors'').

The integration of LLMs provides high flexibility in accessing various system functionalities, reducing the need for users to strictly follow specific command syntax. Instead, users can rely on natural language inputs, which alleviates cognitive load and allows them to focus on their primary tasks.

\noindent\subsection{System Management}
To achieve a coherent coordination between the visualization and interaction functions, we propose a robust system management workflow that integrates three main components.

\noindent\textbf{Spatial Synchronization.} 
The spatial synchronization mechanism maintains continuous alignment between 2D and 3D views, ensuring that interactions in either view are immediately reflected in the other. This consistency is achieved via the bidirectional spatial coherence algorithm, which guarantees accurate transformations between coordinate systems.

\noindent\textbf{Interaction Coordination.} 
Hand gestures and voice commands are processed through a unified command pipeline, which resolves potential conflicts and ensures smooth transitions between interaction modalities. The system also supports hybrid interactions, such as using voice commands to select an anatomical structure while adjusting its visualization parameters with hand gestures. This flexibility empowers users to combine modalities for more efficient and customizable interactions.

\noindent\textbf{State Management.} 
The system employs a centralized state management mechanism based on the MVVM pattern, providing the advantages of (1) real-time synchronization of visualization states across different views; (2) efficient handling of concurrent multi-modal interactions; (3) consistent application of user preferences and visualization parameters; and (4) robust error handling and state recovery mechanisms.


\section{Evaluation}

\subsection{Hypotheses and Study Design}
We conduct a user study and expert interviews to assess our XR system's effectiveness in facilitating medical data visualization and exploration. Our evaluation aims to verify three key hypotheses:

\begin{itemize}[noitemsep,topsep=0pt]
    \item \textbf{H1}: The coordinated 2D-3D visualization will improve users' ability to associate anatomical structures with their corresponding radiological images.
    
    \item \textbf{H2}: The LLM-enabled voice commands will enhance efficiency in volumetric data exploration compared to conventional interface controls.
    
    \item \textbf{H3}: The integrated system will demonstrate improved usability for volumetric medical data visualization, particularly for users with limited medical expertise.
\end{itemize}

\subsection{Experimental Setup}
\noindent\textbf{System Development.}
Our XR system is developed using Unity 2022.3 LTS \cite{unity2022} with Meta Quest 3 headset \cite{metaquest3}. We utilize the 3D-IRCADb-02 dataset~\cite{soler20103d} for volumetric medical data visualization. The system incorporates MRTK3 \cite{mrtk2024} for XR interactions and Meta's Wit.ai \cite{witai2024} for voice recognition. System performance is monitored using Meta Quest Developer Hub \cite{metahorizon2024} and Unity Memory Profiler \cite{unitymemoryprofiler2024}. Given the monitored statistics, our developed system maintains a frame rate of 63-83 FPS with an average memory consumption of 1.27GB, showing stable performances in real-time medical visualization tasks.


\noindent\textbf{User Study.}
We conducted a 15-participant user study, including postgraduate engineering students and junior medical students (13 males, 2 females, aged 23-32), all of whom had limited medical knowledge. We randomly distributed them into three groups: \textbf{G1} (control, using basic UI shown in Fig.~\ref{fig:system_overview}(a)), \textbf{G2} (2D-3D coordination, using enhanced UI shown in Fig.~\ref{fig:system_overview}(b), excluding voice commands), and \textbf{G3} (full features with complete UI in Fig.~\ref{fig:system_overview}(b-d)). After a 10-minute tutorial on how to use our XR system with Meta Quest 3 and covering essential anatomical knowledge for task completion, participants were asked to complete three tasks: 
\begin{itemize}[noitemsep,topsep=0pt]
    \item \textbf{T1}: Spatial Relationship Comprehension Task. Participants were instructed to describe the spatial relationships between multiple anatomical structures.
    \item \textbf{T2}: 2D-3D Correlation Task. Participants were asked to identify specific anatomical structures and describe pathological features in 3D models \& corresponding 2D radiological slices.
    \item \textbf{T3}: Volumetric Data Exploration Task. Participants were given a complex exploration scenario requiring the manipulation of multiple visualization parameters (\emph{e.g.}, opacity, clipping planes) to reveal hidden structures.
\end{itemize}

\noindent\textbf{Expert Interview.}
We conducted semi-structured interviews with three professionals (one senior clinical medicine postgraduate student and two senior hepatobiliary surgeons) to evaluate our system and identified potential improvements. The interviews focused on system usability, educational value, and clinical applicability.

\subsection{User Study Analysis}
  
\noindent\textbf{Task Completion Time.} We collect the data on task completion time to analyze varying operational performances across the three groups (n = 5 per group) in Tab.~\ref{tab:task_time}. Tasks were considered complete only when participants provided correct answers. Incorrect responses were identified as incorrect, prompting participants to retry, but no further feedback or hints were given. For the Spatial Relationship Comprehension Task (\textbf{T1}), the groups (\textbf{G2}: M = 90.4s, \textbf{G3}: M = 92.0s) with 2D-3D coordination feature perform better than the control group (\textbf{G1}: M = 101.8s). For the 2D-3D Correlation Task (\textbf{T2}), the \textbf{G3} group demonstrates the fastest mean completion time (M = 262.4s) compared to the \textbf{G1} (M = 448.4s) and \textbf{G2} (M = 323.4s) groups. For the Volumetric Data Exploration Task (\textbf{T3}), the \textbf{G3} group (M = 78.0s) shows the best performance, followed by \textbf{G2} (M = 82.6s) and \textbf{G1} (M = 99.2s), indicating that voice commands become more effective when using our coordinated 2D-3D visualization. Overall, the \textbf{G3} group achieves the lowest total task time (M = 432.4s), showing consistent and prominent efficiency gains in volumetric medical data exploration tasks.

\noindent\textbf{System Usability Scale.} The System Usability Scale (SUS) provides insights into the perceived usability of our system. Focusing on descriptive statistics, the \textbf{G3} group reports the highest mean SUS score (M) of 87.0 with a standard deviation (SD) of 9.25, followed by the \textbf{G1} group (\emph{i.e.}, the control group, where M = 80.5, SD = 14.83), and the \textbf{G2} group (\emph{i.e.}, the group with the 2D-3D coordination feature only, where M = 74.5, SD = 12.42). These scores suggest a trend towards improved usability with our full-feature XR system, indicated by \textbf{G3}'s mean score falling in the ``excellent'' range as per SUS interpretations~\cite{bangor2009determining}.

\begin{figure}[t]
    \centering
    \includegraphics[width=0.9\linewidth]{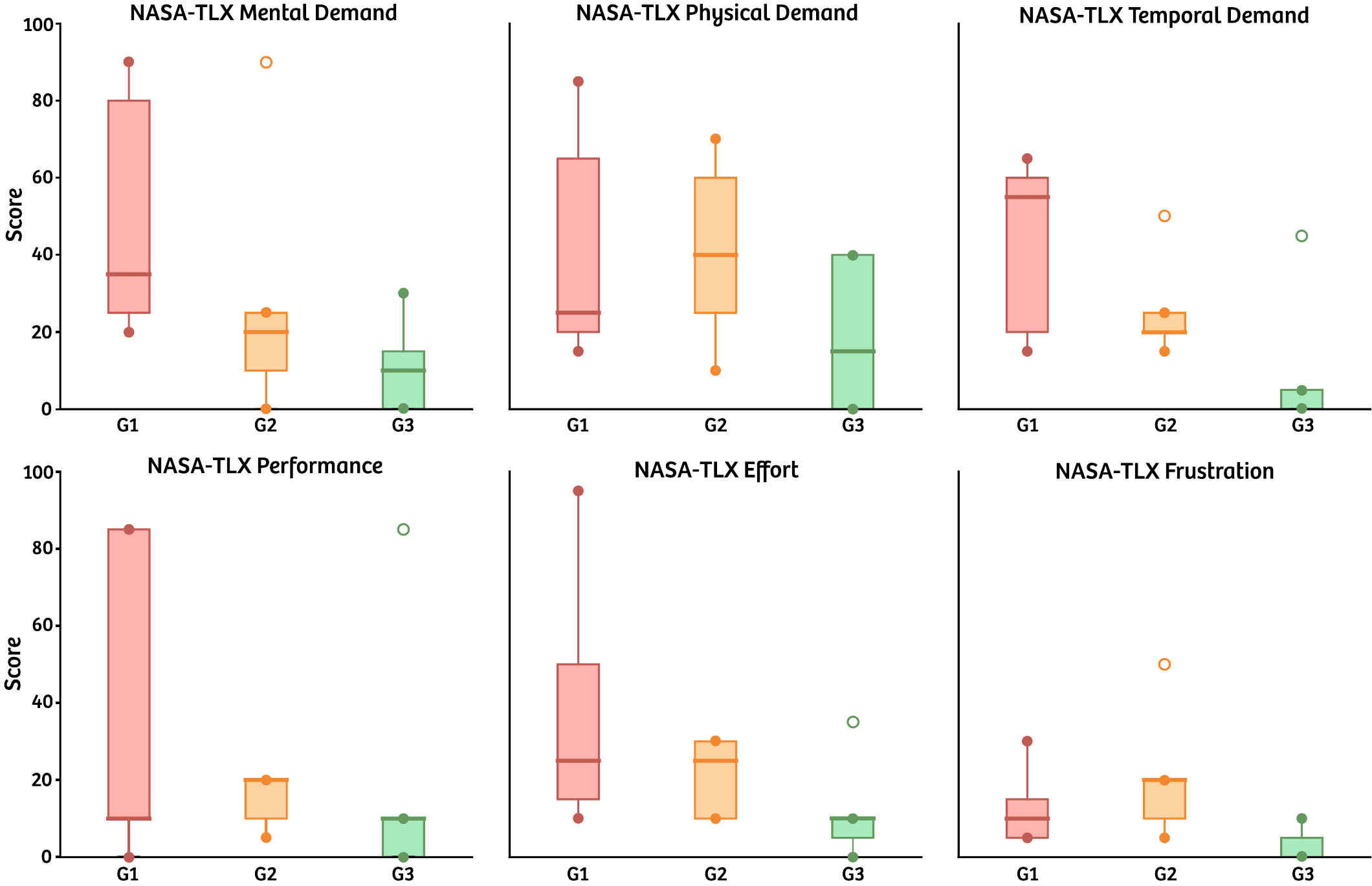}
    \caption{The NASA Task Load Index from user study.
    }
    \label{fig:nasa_tlx}
\end{figure}

\noindent\textbf{NASA-TLX.} 
The NASA Task Load Index for six subjective subscales is analyzed by groups, as shown in Fig.~\ref{fig:nasa_tlx}. In particular, the \textbf{G3} group reports the lowest mean overall workload (M = 12.67, SD = 14.09), followed by \textbf{G2} (M = 25.5, SD = 9.82), and \textbf{G1} (M = 37.5, SD = 21.10). These scores suggest a trend toward reduced cognitive load with our system, particularly when both our proposed 2D-3D coordination and voice command features are enabled. Notably, the experimental data analysis reveals that, compared to the \textbf{G1} group, the \textbf{G3} group requires lower mental demand and temporal demand, while achieves higher perceived performance. 

\begin{table}
\caption{Task Completion Time (Mean (Std), seconds)}
\begin{center}
\begin{tabular}{|c|c|c|c|}
\hline
\multirow{2}{*}{Tasks} & \multicolumn{3}{c|}{User Groups} \\
\cline{2-4}
& \textbf{G1}: \textit{Control} & \textbf{G2}: \textit{2D-3D} & \textbf{G3}: \textit{Full} \\
\hline
\textbf{T1} & 101.8 (40.53) & \textbf{90.4 (45.99)} & 92.0 (27.89) \\
\hline
\textbf{T2} & 448.4 (97.08) & 323.4 (104.21) & \textbf{262.4 (63.03)} \\
\hline
\textbf{T3} & 99.2 (15.75) & 82.6 (13.37) & \textbf{78.0 (14.14)} \\
\hline
Overall & {649.4 (125.70)} & {496.4 (145.73)} & {\textbf{432.4 (89.82)}} \\
\hline
\end{tabular}
\label{tab:task_time}
\end{center}
\end{table}

\subsection{Hypotheses Testing}
The evaluation results support all three hypotheses: \textbf{H1} is validated by \textbf{G2} and \textbf{G3}'s improved performance in spatial comprehension (\textbf{T1}: 10\% faster) and structure identification (\textbf{T2}: 28-41\% faster). \textbf{H2} is confirmed by \textbf{G3}'s superior performance in both structure identification (\textbf{T2}: 18.9\% faster than \textbf{G2}) and volumetric exploration (\textbf{T3}: 5.6\% faster than \textbf{G2}), along with significantly lower cognitive load (NASA-TLX: 50-66\% reduction).
\textbf{H3} is supported by \textbf{G3}'s high usability scores among non-expert users (SUS: 87.0, ``excellent'' range), effective task completion (33.4\% faster than \textbf{G1}), and low cognitive load (NASA-TLX: 66\% reduction vs. \textbf{G1}).

\subsection{Insights from Expert Interview}

The experts highlight several strengths of our system. The coordinated 2D-3D visualization is praised for its effectiveness in supporting spatial understanding, especially in correlating cross-sectional images with 3D anatomical structures. The multi-modal interaction design, especially the LLM-enabled voice commands, is recognized for reducing cognitive load during medical data exploration. The overall system shows particular promise for medical education, with experts noting its potential to help students develop critical spatial comprehension skills essential for anatomical studies.

The experts also identify areas for improvement. The voice recognition system requires further refinement to enhance its accuracy, especially when interpreting complex medical terminology. And the experts suggest incorporating specialized tools for specific clinical applications like surgical planning or patient consultation. These insights provide valuable guidance for future enhancements.

\section{Conclusion}
This paper presents a novel XR system for volumetric medical data visualization, integrating coordinated 2D-3D displays with LLM-enabled multimodal interactions. Preliminary user studies with 15 participants and interviews with 3 experts demonstrate the system’s enhanced usability and functionality, particularly benefiting users with limited medical expertise. While the initial results are promising, areas for improvement are identified for further development. In the future, we will conduct larger-scale and long-term evaluations to validate the system’s effectiveness across different medical specialties and user groups.

\acknowledgments{This work was supported by the Research Grants Council of the Hong Kong Special Administrative Region, China (Project No.: T45-401/22-N); and in part by The Chinese University of Hong Kong (Project No.: 4055212).}

\bibliographystyle{abbrv-doi}

\bibliography{references}
\end{document}